# STUDY ON THE FRAGMENTATION OF SHELLS

Falk K. Wittel[1], Ferenc Kun[2], Bernd H. Kröplin[1] and Hans J. Herrmann[3]

[1]*Institute for Statics and Dynamics of Aerospace Structures, University of Stuttgart, Pfaffenwaldring 27, D-70569 Stuttgart, Germany*
[2]*Department of Theoretical Physics, University of Debrecen, P.O. Box:5, H-4010 Debrecen, Hungary, e-mail: feri@dtp.atomki.hu*
[3]*Institute for Computer Physics, University of Stuttgart, Pfaffenwaldring 27, D-70569 Stuttgart, Germany*

**Abstract.** Fragmentation can be observed in nature and in everyday life on a wide range of length scales and for all kinds of technical applications. Most studies on dynamic failure focus on the behaviour of bulk systems in one, two and three dimensions under impact and explosive loading, showing universal power law behaviour of fragment size distribution. However, hardly any studies have been devoted to fragmentation of shells. We present a detailed experimental and theoretical study on the fragmentation of closed thin shells of various materials, due to an excess load inside the system and impact with a hard wall. Characteristic fragmentation mechanisms are identified by means of a high speed camera and fragment shapes and mass distributions are evaluated. Theoretical rationalisation is given by means of stochastic break-up models and large-scale discrete element simulations with spherical shell systems under different extreme loading situations. By this we explain fragment shapes and distributions and prove a power law for the fragment mass distribution. Satisfactory agreement between experimental findings and numerical predictions of the exponents of the power laws for the fragment shapes is obtained.

## 1. Introduction

Closed shells made of solid materials are used in every day life, in industrial applications in form of containers, pressure vessels or combustion chambers and also in nature *e.g.*, as nature's oldest container for protecting life - the egg-shell. From a structural point of view, aircraft vehicles, launch vehicles like rockets and building blocks of a space station are also shell-like systems. These constructions can fail due to an excess load which can arise by slowly driving the system above its stability limit. However, imparting high energy in a short time, like in impact cases or for pressure pulses, due to an explosive shock inside the shell, result in their fragmentation. The explosive fragmentation of a rocket tank resulting in space debris is such a case, endangering other space activities for years onward with an enormous social impact because of the human and financial costs arising as a consequence of accidents [1].

Fragmentation, in other words the breaking of particulate materials into smaller pieces, is abundant in nature occurring on a broad range of length scales from meteor impacts through geological phenomena and industrial applications down to the break-up of large molecules and heavy nuclei [2-5]. The most striking observation concerning fragmentation is that the distribution of fragment sizes shows power law behaviour independent of the way of imparting energy, relevant microscopic interactions and length scales involved, with an exponent depending solely on the dimensionality of the system [3-13]. Detailed experimental and theoretical studies have revealed that universality prevails for large enough input energies when the system fragments into small enough pieces [3-7], however, for lower energies a systematic dependence of the exponent on the input energy is evident [11]. Recent investigations on the low energy limit of fragmentation suggest that the power law distribution of fragment sizes arises due to an underlying critical point [7-10]. Former studies on fragmentation have focused on bulk systems in one, two and three dimensions; however, hardly any studies are devoted to the fragmentation of shells [12-14]. The peculiarity of the fragmentation of closed shells originates from the fact that their local structure is inherently two-dimensional, however, the dynamics of the systems, deformation and stress states are three dimensional which allows for a rich variety of failure modes.

## 2 Experiments on Shell Fragmentation
### 2.1 Impact and Explosion of Shell Systems

To characterize the break-up process and fragmentation mechanisms of shells, explosion and impact experiments have been performed on shells made of various materials. Ordinary white and brown hen eggs and quail eggs were emptied through two holes, washed, rinsed and dried to get thin brittle shells of disordered brittle bio-ceramic [15]. For the other extreme of amorphous material, hollow glass spheres were used. Impact experiments were performed by catapulting the object with a set of rubber bands, against a hard wall on the bottom of a soft plastic bag, prohibiting secondary fragmentation and assuring that no fragments were lost for further evaluation (Fig. 2). Explosions were realized by filling the objects with detonating gas and electric ignition inside a large, soft plastic bag as well. To study fragmentation mechanisms, the break-up process during the explosion was followed by means of a Photron APX ultra high speed camera with frame rate 15000/s (see Fig. 1).

From the high-speed pictures we can distinguish two different breaking regimes for in-plane and out-of-plane deformation. The analysis of egg-shells has shown that the in-plane break-up starts with the nucleation of a few cracks at the flatter end of the egg. Since the energy stored in the expanding system at the instant of crack nucleation is high compared to the energy released by the free fracture surface, cracks propagate at high speeds. The instability of the propagating cracks results in sequential crack bifurcations at almost regular distances (Fig.1 IA-D). The propagating sub-branches accelerate due to the further expansion of the shell and can undergo further splitting, giving rise to a hierarchical tree-like crack pattern (Fig.1 IB). Fragments are formed along the main cracks by the merging of adjacent side branches at almost right angles (Fig.1 IC). In the later stages of the fragmentation, the out-of-plane deformation dominates, giving rise to further cracking of elongated fragments (Fig.1 ID). The out-of-plane cracking can again proceed in a sequential manner, typically breaking fragments into two pieces (binary break-up), until a stable configuration is reached.

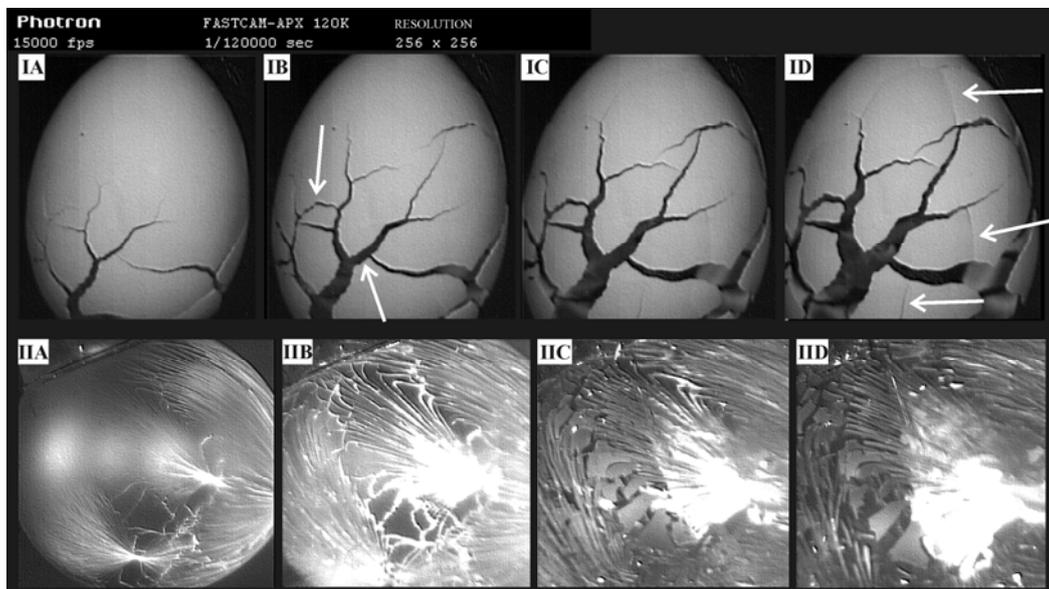

*Figure 1*: High speed pictures of exploding egg shell (I) and hollow glass sphere (II). Fragment creation by branching and merging of the hierarchical crack-tree (IB) and cracks leading to binary break-up of fragments (ID) are indicated.

The described fragmentation mechanisms should be generic for materials with a highly disordered microstructure. However, for shells made of amorphous materials like glass, mechanisms differ significantly. For explosions, the break-up is observed to start at some "hot spots" with random positions on the surface (see Fig.1 IIA), from which long straight cracks radiate without any apparent branching. This mechanism results in a large number of long thin fragments, making them unstable against out-of-plane bending and sequential breaking, respectively. Fragments formed by impact with a hard wall are comparable.

## 2.2 Fragment Shape Analysis

The resulting pieces are carefully collected and placed on the tray of a scanner without overlap. In impact experiments a few thousand fragments are generated, while explosions give rise to a few hundred fragments. In the digitalized images fragments occur as, for instance, black spots on a white background and are further analyzed by a cluster searching code (see Fig.2). The mass $m$ and fracture surface $A$ of fragments were determined as the number $N$ of pixels of pieces in the scanned image and the contour length of the spots. Shattered fragments of sizes comparable to normal dust pieces in the air were excluded from the analysis by setting the cut-off size to a few pixels. Finally, the analysed fragment mass is over 97% of the total mass of the test objects for all experiments.

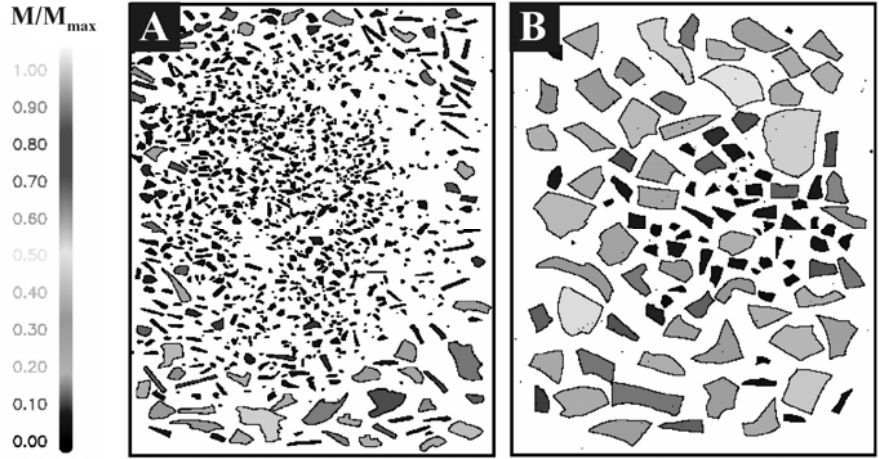

*Figure 2:* Digitalized shell fragments of different materials obtained in explosion: (A) glass and (B) egg-shell. Grey values represent fragment mass $M$ normalized by the largest fragment mass $M_{max}$.

It can be observed, that fragments are compact, two-dimensional objects with low fracture surface roughness for all materials. However, their overall shape can strongly vary from isotropic (egg-shell, Fig.2B) to highly anisotropic (glass, Fig.2A) depending on the fragmentation mechanism. The linear extension of fragments is characterized by their radius of gyration $R_g$ to $R_g^2 = (1/N) \sum_{i \neq j=1}^{N} (\vec{r}_i - \vec{r}_j)^2$, where the sum goes over the pixels $r_i$ of the fragments. The average mass $<m>$ over $R_g$ (Fig.3A) reveals how the shape of fragments varies with their size for different fragmentation experiments. In all cases high quality power law functional forms $<m> \sim R_g^\alpha$ are obtained with the exponent $\alpha$ depending on the structure of the crack pattern. Since egg-shell fragments are of regular isotropic shape, their mass increases quadratically with $R_g$ and $\alpha = 2 \pm 0.05$ is obtained (see Fig.3A). Needle-like glass fragments are very elongated and have a significantly lower exponent $\alpha = 1.5 \pm 0.05$, while small glass fragments show a crossover to an isotropic shape. The value of $\alpha < 2$ implies that fragments have self-affine character, meaning that the larger they are, the more elongated they get.

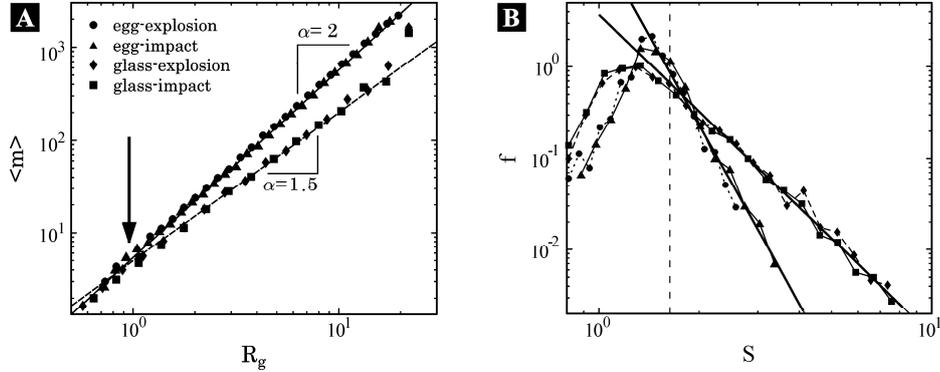

*Figure 3:* Function $<m>$ over $R_g$ with crossover point of glass fragments (A) and shape distribution of fragments (B) with identical legend as (A).

The shape distribution depends on the definition of shape parameters. A classical characterization of the shape of fragments is done by the fracture surface to mass ratio $A/m$. A more transparent characterization of fragment shapes can be obtained by introducing a dimensionless shape parameter $S=R_g \cdot A/m$. For rectangular fragments $S = (a+b)\sqrt{a^2+b^2}/(\sqrt{3}ab)$ and isotropic fragments with $a \approx b$ leads to $S \approx 1.63$ indicated by the line in Fig.3B. If $a >> b$ fragments are elongated and $S \approx a/b$. Fragments of a low degree of anisotropy, irrespective of their size, contribute to the maximum of $f(S)$ in the vicinity of $S \approx 1.63$. Since egg-shell fragments are mostly isotropic in all sizes, the distribution $f(S)$ decreases rapidly over a narrow interval of $S$ with a power law decay $f(S) \sim S^\beta$. When cracking mechanisms favour the formation of anisotropic fragments like for glass, $\beta = 3.5 \pm 0.2$ is obtained.

**2.3 Fragment Mass Analysis**

Due to the violent nature of fragmentation processes, the fragment masses as shown in Fig. 4 are one of the few quantitative measures. For the impact experiment, a power law behaviour of the distribution $F(m) \sim m^{-\tau}$ can be observed over three orders of magnitude where the value of the exponent can be determined as $\tau = 1.35 \pm 0.02$ with high precision. Explosion experiments also result in a power law distribution of the same value of $\tau$ for small fragments with a relatively broad cut-off for the large ones. Note that the exponent $\tau$ is significantly different from the experimental and theoretical results when fragmenting two-dimensional bulk systems where $1.5 \leq \tau \leq 2$ has been found [3-10].

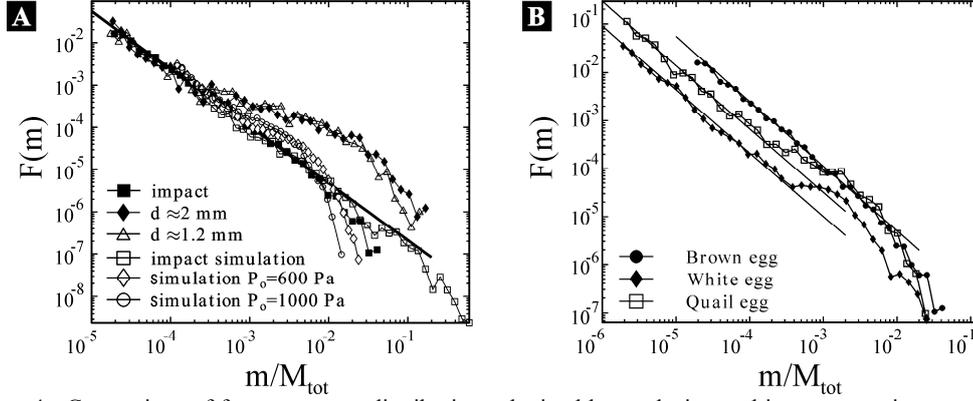
Figure 4: Comparison of fragment mass distributions obtained by explosion and impact experiments of egg shells with different hole sizes (A) and impact for different materials (B) averaging over 10-20 realizations for each curve.

## 3 Simulation of the Break-up Process
To rationalize the experimental findings, we use two different approaches, discrete element methods (DEM) and stochastic break-up models. DEM are capable of simulating shell fragmentation of highly disordered brittle materials rather well and they provide isotropic fragment shapes in agreement with experiments, but have difficulties to capture mechanisms resulting in long straight cracks. For this purpose stochastic binary break-up models based on models by [2,16] are utilized.

### 3.1 Predictions for Fragment Shapes
First we assume that all shell fragments are of rectangular shape with the side lengths $a$ and $b$. The fracture surface $A$, mass $m$ and radius of gyration $R_g$ simply follows as $A = 2(a+b)$, $m=ab$ and $R_g = \sqrt{a^2+b^2}/(2\sqrt{3})$. Using the aspect ratio $r=a/b$, the fragment mass can be expressed as $m \sim R_g^2/(r+1/r)$, where even for moderately elongated fragments the approximation $m \sim R_g^2/r$ is valid. For fragments with exponents $\alpha < 2$, $r$ must increase as a power of $R_g$ so that $r \sim R_g^\delta$. Consequently, $m \sim R_g^{2-\delta}$ and $\alpha = 2-\delta$ follows, with an experimental value of $\delta = 1/2$. Note that self-affinity of fragment shapes and anisotropy is a peculiarity of shell fragmentation since for $d$-dimensional bulk fragmentation (d=2,3) $\alpha = d$ and $\delta = 0$ is observed.

The model represents the primary out-of-plane cracking mechanism of the shell by assuming binary break-up of rectangular shaped fragments with continuous mass distribution at a fixed initial aspect ratio $r$ (see Fig.5A). By setting $r$ from isotropic ($r \approx 1$) to an anisotropic needle-like shape ($r \gg 1$) the effect of the material-dependent primary cracking mechanism can be considered. The fragments are assumed to undergo a sequential binary break-up process, where they break into two pieces of equal mass with a probability $p \leq 1$ at each step of the hierarchy. Consequently, fragments have a $1-p$ chance to keep their actual size and survive until the final state. Note that $p$ is assumed to be completely generation independent for simplification. We choose a side of a rectangle to break with a probability proportional to its length. Starting from an uniform distribution of fragment masses, the binary break-up mechanism gives rise to a power law mass distribution $F(m) \sim m^{-\tau}$ where the initial form of the distribution mainly affects the shape of the cut-off in the final state. The exponent $\tau = 2 + \ln p / \ln 2$ depends on the parameter $p$ of the model. Physically meaningful values are obtained in the range $1/4 \leq p \leq 1$, with a resulting $\tau$ of $0 \leq \tau \leq 2$, but the actual value of $p$ has only a minor effect on the fragment shapes, justifying our assumption. Computer simulations of the model were made, varying the initial aspect ratio $r$ to model different materials for fixed $p$. The hierarchical process was followed for up to 30 generations resulting in up to $10^7$ fragments.

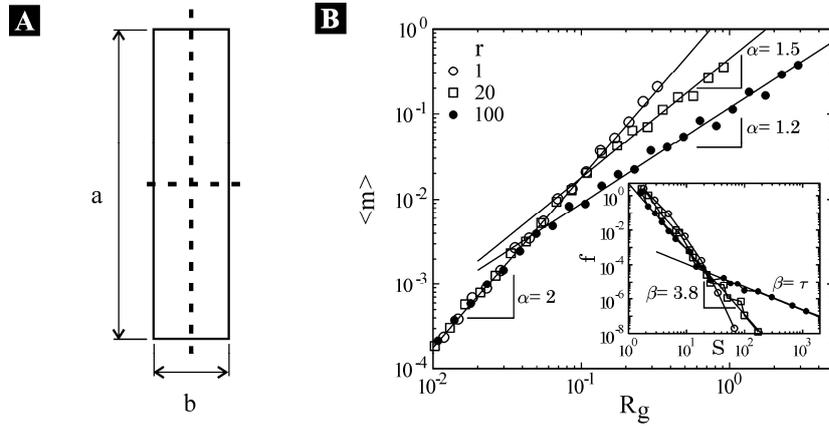

*Figure 5:* Binary break-up model (A) resulting functions (B) and shape parameter distributions (inset of B).

Evaluating $m$, $R_g$ and $S$, we show in Fig. 5B that for all values of $r$, the average fragment mass exhibits a power law dependence on the radius of gyration $<m> \sim R_g^\alpha$ as observed in the experiments. For initially isotropic fragments ($r \approx 1$) $\alpha = 2 \pm 0.05$, since they remain mainly isotropic at all levels of the hierarchy as can be observed for egg shells. For glass fragments with high anisotropy ($r \gg 1$), cracks occur mostly on the long side of the rectangle with a value of $1 \leq \alpha \leq 2$, reducing $r$ until fragments become isotropic again (see Fig. 5B). For glass spheres values of $r \approx 20$ are realistic, and values of $\alpha = 1.5 \pm 0.05$ are obtained in agreement with experimental findings. The hierarchical break-up process also leads to a power law distribution of the shape parameter $f(S) \sim S^{-\beta}$. For highly elongated fragments $\beta \to \tau$ (see Fig. 5B*inset*) while for smaller fragments a crossover to a second power law regime with a higher exponent of $\beta = 3.8 \pm 0.3$ is observed. Initially moderately elongated fragments ($r \approx 20$) only show the higher exponent while for isotropic fragments *f(S)* decreases rapidly over a relatively narrow range in satisfactory agreement with experiments.

**3.2 Scaling Behaviour of Fragment Masses**
To consider the dynamics and evolution of the fragmentation process we have worked out a three dimensional discrete element model of spherical shells by discretizing the surface of a unit sphere into randomly shaped triangles (Delaunay triangulation) and Voronoi polygons. For simplicity, our theoretical study is restricted to spherical shells. The nodes of the triangulation represent point-like material elements whose mass is defined by the area of the Voronoi polygon assigned to it (see Fig.6A). The nodes are connected by springs which behave linear-elastically up to failure. Therefore we obtain a three dimensional central force random network, whose disorder is introduced solely by the randomness of the tessellation. Hence, the mass of the nodes, the length and cross-section of the springs are determined by the tessellation (quenched structural disorder). After prescribing the initial conditions of a specific fragmentation process studied, the time evolution of the system is followed by simultaneously solving the equation of motion of all nodes. In order to account for crack formation in the model springs are assumed to break during the time evolution of the system when their deformation $\varepsilon$ exceeds their breaking threshold $\varepsilon_c$, resulting in a random sequence of breakings, due to the disordered spring properties. As a result of successive spring breakings micro cracks nucleate, grow and merge on the spherical surface giving rise to a complete break-up of the shell into fragments (see Fig.6B). The process is stopped when the system has reached a relaxed state. The typical model consists of ~23000 nodes with ~120000 spring elements. Due to the nature of this model, the smallest possible crack is sharply defined by the characteristic length scale of the model, giving the possibility to simulate simultaneous dynamic growth of many interacting crack. The model is described in more detail in [12,13].

In computer simulations two different ways of loading have been realized to model the experimental conditions, representing limiting cases of energy input rates: *(i) pressure pulse* and *(ii) impact* load starting from an initially stress free state. A *pressure pulse* in a shell is carried out so that a fixed internal pressure $P_0$ is imposed giving rise to an expansion of the system with a continuous increase of the imparted energy $E_{tot}=P_0 \cdot \Delta V$, where $\Delta V$ denotes the volume change with respect to the initial volume $V_0$. Since the force *F* acting on the shell is proportional to the actual surface area *A*, the system is driven by an increasing force $F \approx P_0 \cdot A$ during the expansion process. The impact loading represents the limiting case of in-stantaneous energy input $E_{tot}=E_0$. This is considered in the simulation by giving a fixed radial outwardly directed initial velocity $v_0$ to the nodes. Simulations with broadly varied control parameters $P_0$ and $E_0$ have revealed a substantial difference between the two ways of fragmentation. For both ways of loading a rather uniform stress and deformation state arises so that during the expansion process first overstrained springs break in an uncorrelated manner generating micro cracks on the surface.

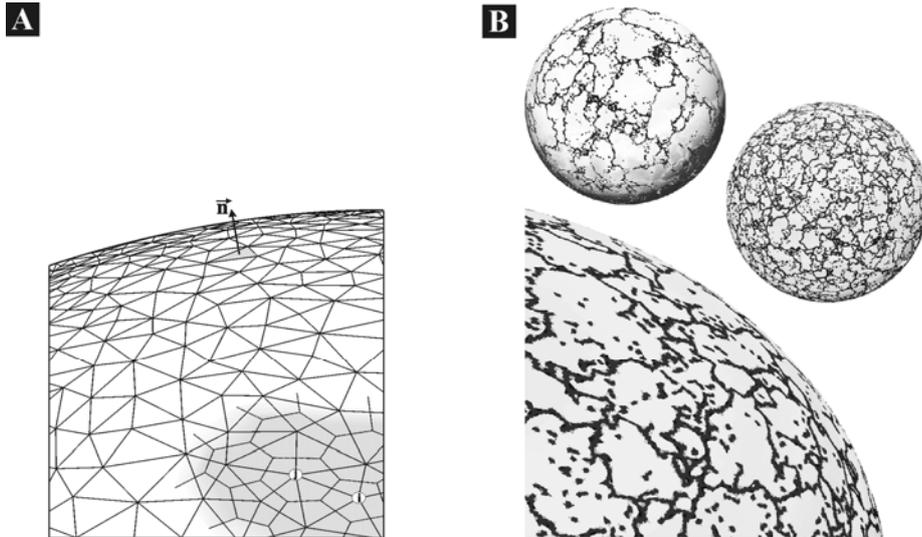

*Figure 6:* (A) Delaunay triangulation of a spherical surface with dual Voronoi lattice and (B) cracks on the shell surface for impact energies $E_0/E_c \approx 2.8$ (left and closeup) and pressure loading with $P_0/P_c \approx 4.0$ (right), projected on initial configuration.

When $P_0$ exceeds the critical pressure $P_c$, the expanding shell surpasses a critical volume $V_c$ at which fragmentation sets in, *i.e.* abruptly a large amount of springs break rapidly forming cracks which grow and join resulting in shell pieces surrounded by a free crack surface (fragment). First large fragments are formed which then break-up into smaller pieces until the surviving springs can sustain the remaining stress, see Fig. 6B. The transition from the damaged state with the shell keeping its integrity to the fragmented state where the system disintegrates into pieces occurs abruptly at $P_c$. Note that fragmentation due to out of plane bending occurs due to the curvature of fragments. Under *impact loading* however, the initial radial velocity implies a prescribed path for the motion of material elements resembling the strain-controlled loading of bulk specimens. Similar to the pressure loading case, simulations revealed that a critical value of the imparted energy $E_c$ can be identified. Below $E_c$ the shell maintains its integrity, suffering only damage, while exceeding $E_c$ gives rise to a complete fragmentation of the shell (see Fig. 6B).

A quantitative characterization of the break-up process with increasing control parameters is possible by monitoring the average mass of the largest fragment normalized by the total mass $<M_{max}/M_{tot}>$. $M_{max}$ is a monotonically decreasing function of both $P_0$ and $E_0$, however, the functional forms differ (see Fig. 7A-B). At low pressure values in Fig. 7A $M_{max}$ is practically equal to the total mass since hardly any fragments are formed. Above $P_c$ however, $M_{max}$ gets significantly smaller than $M_{tot}$, indicating the disintegration of the shell into pieces.

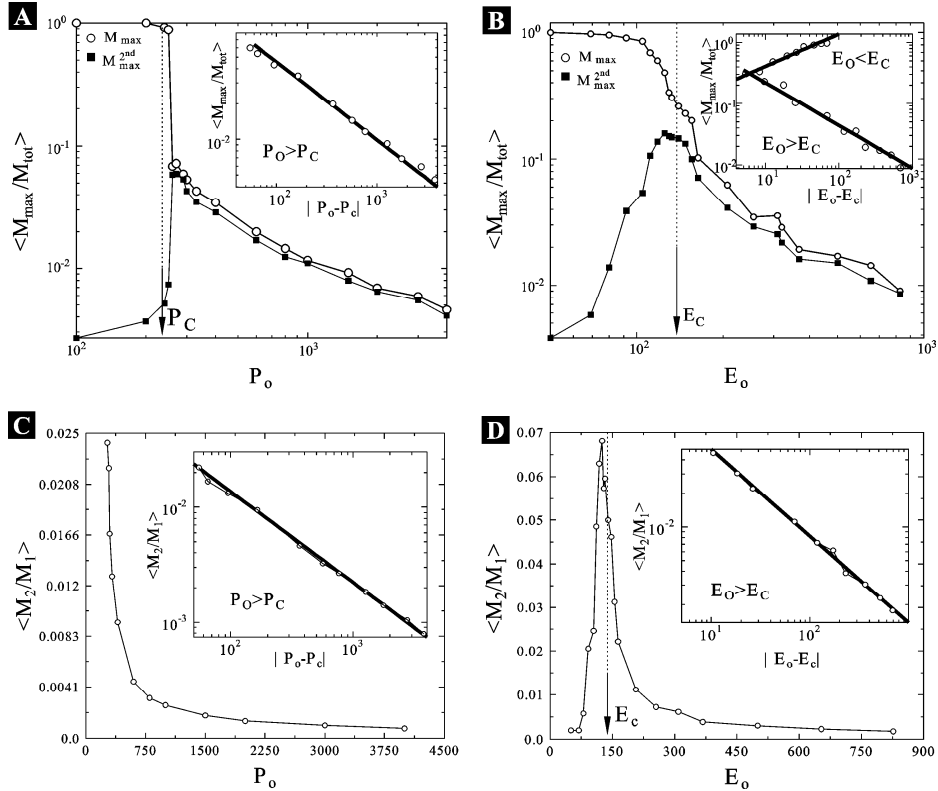

*Figure 7:* Average mass of largest and second largest fragments (A,B) normalized by the total mass as a function of the imposed pressure $P_0$ (A) and impact energy $E_0$ (B). The average fragment mass as functions of control parameters $P_0$ and $E_0$ is shown in (C,D). Insets present the results as a function of the distance from the critical point $P_c$ and $E_c$.

The value of the critical pressure $P_c$ needed to achieve fragmentation and the functional form of the curve above $P_c$ were determined by plotting $<M_{max}/M_{tot}>$ as a function of the difference $|P_0 - P_c|$ varying $P_c$ until a straight line is obtained on a double logarithmic plot. The power law dependence on the distance from the critical point $<M_{max}/M_{tot}> \approx |P_0 - P_c|^{-\alpha}$ for $P_0 > P_c$ is evidenced by the inset of Fig. 7A, where $\alpha=0.66\pm0.02$ was determined showing good agreement with analytic predictions [11,12]. Fig. 7A demonstrates the abrupt nature of the transition between the two regimes for pressure loading at a well defined critical pressure $P_c$ below which practically no fragments are formed. The shell cracks but all the cracks arrest and the shell retains its integrity. Small fine powder can be broken out of the shell along the cracks but still the mass of the largest fragment is practically equal to the total mass. The quantity $M_{max}/M_{tot}$ in our study plays the role of the order parameter similar to the spanning cluster in percolation.

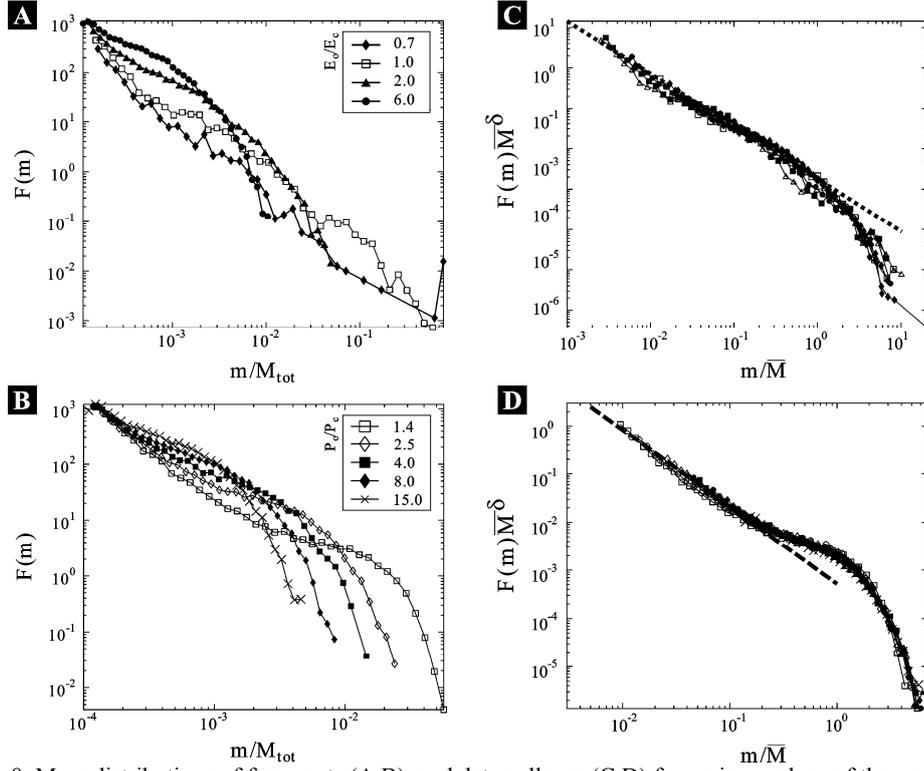

*Figure 8:* Mass distributions of fragments (A,B) and data collapse (C,D) for various values of the control parameter $E_0$ and $P_0$.

For impact loading $<M_{max}/M_{tot}>$ proved to be a continuous function of $E_0$, however, it also shows the existence of two regimes of the break-up process with a transition at a critical energy $E_c$. In the inset of Fig. 7B $<M_{max}/M_{tot}>$ is shown as a function of the distance from the critical point $|E_0 - E_c|$, where $E_c$ was determined numerically in the same way as $P_c$. Contrary to the pressure loading, $<M_{max}/M_{tot}>$ exhibits a power law behaviour on both sides of the critical point but with different exponents $<M_{max}/M_{tot}> \approx |E_0 - E_c|^\beta$ for $E_0 < E_c$ and $<M_{max}/M_{tot}> \approx |E_0 - E_c|^{-\alpha}$ for $E_0 > E_c$ where the exponents were obtained as $\beta = 0.5 \pm 0.02$ and $\alpha = 0.66 \pm 0.02$. Note that the value of $\alpha$ coincides with the corresponding exponent of the pressure loading.

We also evaluated the average fragment mass $<M>$ as the average value of the ratio of the second $M_2$ and first $M_1$ moments of fragment masses [8]. The behaviour of $<M>$ again clearly shows the existence of two regimes of the break-up process with a transition at the critical points $P_c$ and $E_c$ (see Fig. 7C-D). Under pressure loading due to the abrupt disintegration $<M>$ can only be evaluated above the critical point $P_c$, while for the impact case $<M>$ has a maximum at the critical energy $E_c$ typical for continuous phase transitions. In both loading cases $<M>$ has a power law dependence on the distance from the critical point, i.e. $<M> \sim |E_0-E_c|^{-\gamma}$, and $<M> \sim |P_0-P_c|^{-\gamma}$ hold with the same value of the exponent $\gamma= 0.8\pm0.02$ (inset of Fig. 7C-D).

An important characteristic quantity of our system is the mass distribution of fragments $F(m)$. For *impact loading* representative examples are shown at an energy value below, above, and close to $E_c$ (see Fig. 8A). For *pressure loading* $F(m)$ can only be evaluated above $P_c$ (see Fig. 8C). At energies $E_0<E_c$ it can be observed that $F(m)$ has a pronounced peak for large fragments due to the presence of large damaged pieces. Approaching the critical point $E_c$, the peak gradually disappears and the distribution becomes a power law at $E_c$. Above the critical point the power law remains for small fragments, followed by an exponential cut-off for the large ones. Note the agreement of functional forms of $F(m)$ in Figs. 8A,C with the experimental findings on the impact and explosion of eggs (Fig. 4A). Figs. 8B,D demonstrate that rescaling the distributions above the critical point by plotting $F(m)<M>^\delta$ as a function of $m/<M>$ an excellent data collapse is obtained with $\delta=1.6\pm0.03$. The data collapse implies the validity of the typical scaling form $F(m) \sim m^{-\tau} \cdot f(m/<M>)$ for critical phenomena. The cut-off function $f$ has a simple exponential form $exp(-m/<M>)$ for impact loading, and a more complex one also containing an exponential component for the pressure case. The rescaled plots enable an accurate determination of the exponent $\tau$, where $\tau=1.35\pm0.03$ and $\tau=1.55\pm0.03$ were obtained for impact and pressure loading, respectively. Beside the qualitative agreement of the distributions with the experimental results on egg fragmentation, a good quantitative agreement of the exponent $\tau$ is evidenced between simulations and the impact loading of shells (see Fig. 4A).

## 4 Conclusions

We have presented a theoretical and experimental study of the fragmentation of closed brittle shells arising due to an excess load. We have performed experiments on explosion and impact fragmentation of various shells, resulting in power laws for fragment shapes, shape and fragment size distributions. Stochastic binary break-up models and discrete element models rationalize these observations. We have proven experimentally and in simulations, that the fragmentation of shells is different from other bulk fragmentation processes since:

- Shells are locally seen 2D-objects, with 3D-dynamics leading to additional failure modes like the out-of-plane bending failure.
- Scaling laws of fragment shapes with exponents $\delta$ and $\beta$ are robust features of shell fragments, which are not observed for bulk fragmentation.
- The exponent for power law fragment mass distributions $\tau$ also differs significantly from that of the two and three dimensional bulk systems giving rise to the educated guess that the fragmentation of closed shells belongs to a new universality class.

**Acknowledgement**

The authors are grateful to K.J. Måloy from the University of Oslo for the possibility to use the laboratory and to H. Klinkrad from ESA for valuable discussion. This work was supported by the DFG project SFB381. F. Kun was supported by OTKA T049209, M041537 and by the Gy. Békési Foundation of HAS.